\documentclass[conference]{IEEEtran}
\IEEEoverridecommandlockouts

\usepackage{cite}
\usepackage{amsmath,amssymb,amsfonts}
\usepackage{algorithm} 
\usepackage{algorithmic}
\usepackage{graphicx}
\usepackage{textcomp}
\usepackage{xcolor}
\def\BibTeX{{\rm B\kern-.05em{\sc i\kern-.025em b}\kern-.08em
    T\kern-.1667em\lower.7ex\hbox{E}\kern-.125emX}}

\usepackage{hyperref}
\usepackage{subcaption} 

\begin{document}

\title{Frenzy: A Memory-Aware Serverless LLM Training System for Heterogeneous GPU Clusters \\

}

 \author{\IEEEauthorblockN{1\textsuperscript{st} Zihan Chang}
 \and
 \IEEEauthorblockN{1\textsuperscript{st} Sheng Xiao}
 \and
 \IEEEauthorblockN{2\textsuperscript{nd} Shuibing He}
 \and
 \IEEEauthorblockN{3\textsuperscript{rd} Siling Yang}
 \and
 \IEEEauthorblockN{4\textsuperscript{th} Zhe Pan}
 \and
 \IEEEauthorblockN{5\textsuperscript{th} Dong Li}
 }

\maketitle

\begin{abstract}
Utilizing heterogeneous GPU clusters for training large language models (LLMs) is becoming a prevalent trend in the future. 
Existing work only effective on a given number of GPUs, often neglecting the complexities involved in manually determining the specific types and quantities of GPUs needed, which can be a significant burden for developers.
To address this issue, we propose Frenzy, a memory-aware serverless computing method for heterogeneous GPU clusters. Frenzy allows users to submit models without worrying about underlying hardware resources. First, Frenzy predicts the required number and type of GPUs by estimating the GPU memory usage of the LLM. Then, it employs a low-overhead heterogeneity-aware scheduling method to optimize training efficiency. We validated Frenzy's performance by conducting multi-task LLM training tests on a heterogeneous GPU cluster with three different GPU types. The results show that Frenzy's memory usage prediction accuracy exceeds 92\%, the scheduling overhead is reduced by 10 times, and it reduces the average job completion time by 12\% to 18\% compared to state-of-the-art methods.
\end{abstract}


\section{Introduction}
Training large language models (LLMs) is of paramount importance in modern artificial intelligence research. Traditionally, the training of such models has been conducted on homogeneous GPU clusters, where all GPUs share the same specifications \cite{Alpa, Pollux, Gandiva}. However, using traditional methods often leads to inefficient utilization of different GPUs, resulting in underutilization of some GPUs \cite{mlaas}. Moreover, with the rapid iterative updating of GPUs, many clusters of enterprises, private organizations, or individuals often have different types of GPUs \cite{LagSca-Clus-Ana}. To address these issues, recent efforts have focused on leveraging heterogeneous GPU clusters for training, which can better utilize the diverse computational and memory capabilities of different GPU types \cite{Heterogeneity-Aware,Metis,Sia}. 


Existing work which aims at optimizing LLM training on heterogeneous GPU clusters \cite{Sia}, typically relies on relatively complex algorithms (e.g., Integer Linear Programming), to automatically set up parallel training and task scheduling given a specific number of GPU cards. However, these approaches have two significant issues.

The main issue associated with manually selecting GPUs for training LLMs on heterogeneous GPU clusters are the heavy workload for developers and the risk of errors. Developers are tasked with the intricate job of accounting for GPU heterogeneity, available resources, and model size when determining the types and quantities of GPUs to deploy, which constitutes a considerable burden.
Moreover, manually specifying the number of GPUs can lead to errors. Insufficient allocation may cause out-of-memory (OOM) errors during training, while over-allocation can result in low GPU utilization and reduced training efficiency. 

Another issue is the excessive scheduling time. Scheduling in a heterogeneous cluster involves considering numerous factors, such as heterogeneous computational power, model training time, the number of tasks, and available resources. Using complex algorithms like Integer Linear Programming (ILP) can lead to a large search space, resulting in excessively long scheduling times. 

To address these issues, we propose Frenzy, a memory-aware serverless computing scheduling system. 
First, Frenzy estimates GPU peak memory usage for LLMs and determines the appropriate GPU types and the number of GPUs required for training. Upon submission of a model training task by the user, Frenzy incorporates the designated model parallelism strategy along with the provided model parameters to estimate the model states and activations anticipated during the training of LLMs. This estimation facilitates the prediction of the peak GPU memory consumption that the model is likely to incur. Frenzy then generates resource requirement plans, from which the specific number of GPU cards needed for various types of GPUs can be determined. 

Additionally, we design a scheduling system for heterogeneous GPU clusters that combines the estimated GPU requirements to achieve efficient and low-overhead task-to-cluster scheduling. By leveraging the characteristics of heterogeneous clusters and the efficiency of bin-packing algorithms, Frenzy achieves both low overhead and highly effective online scheduling. Frenzy abstracts the underlying hardware details from the users, automating the deployment and training of user-submitted models without requiring them to manage the hardware resources manually. This ensures a seamless and user-friendly experience, while optimizing resource utilization and training efficiency. 


In summary,  the main contributions of Frenzy is as follows:
\begin{itemize}
    \item We propose Frenzy, which is the first system introducing serverless computing to heterogeneous clusters, significantly reducing the burden on developers.

    \item We designed a memory-aware resources predictor (MARP) to accurately predict the peak memory usage during LLM training, enabling the estimation of the required number of different types of GPUs.
\end{itemize}

\begin{itemize}
    \item We developed a low-overhead heterogeneity-aware scheduling (HAS) strategy that efficiently schedules tasks based on the resource requirements predicted by MARP.
\end{itemize}

\begin{itemize}
    \item We validated the accuracy of Frenzy's memory prediction using Megatron and conducted experiments on both real heterogeneous clusters and simulator-based heterogeneous clusters. The results show that Frenzy's memory prediction accuracy ranges from 92\% to 98\%, scheduling overhead reduced 10 times, and the average job completion time is reduced by 12\%-18\% compared to SOTA works.
\end{itemize}

\section{Background }

\subsection{Serverless Computing}\label{AA}


Serverless computing, or Function-as-a-Service (FaaS), abstracts the underlying infrastructure, allowing developers to focus on writing codes without managing servers or specifying resource allocations\cite{ElasticFlow}. For training large language models (LLMs), serverless platforms automatically provision and scale the necessary computational resources, including GPUs, based on workload demands. This eliminates the need for manual resource specification, simplifying setup and management. By dynamically allocating resources, serverless computing ensures optimal performance and cost-efficiency. These platforms use sophisticated orchestration and scheduling algorithms to distribute workloads across various GPU types, enhancing flexibility and resource utilization. They also support elastic scaling, handling sudden workload increases by provisioning additional resources on-the-fly. This automation and flexibility make serverless computing a powerful tool for accelerating the development and deployment of LLMs, reducing both technical and operational complexities.

\subsection{Heterogeneous training}

Heterogeneous computing involves using a combination of different types of processors and accelerators within a single system to perform computational tasks. In the context of training LLMs, this approach allows each GPU type to handle tasks that best match its capabilities, leading to improved performance and resource utilization, and it allows organizations to leverage existing hardware resources and avoid the need for uniform, high-end GPUs across the entire system. 

Different GPU types vary in memory capacity and computational capabilities, impacting their ability to handle large models and datasets. High-end GPUs like the NVIDIA A100 \cite{a100} have up to 80 GB or 40 GB of memory, making them suitable for training very large models, while mid-range GPUs like the RTX 3090 \cite{3090} have 24 GB of memory, which is substantial but may require more frequent data offloading for larger models. Higher memory capacity allows for more efficient training with larger batches and more model parameters, whereas lower-capacity GPUs may introduce overhead with smaller batch sizes or more frequent data exchanges. Different link bandwidths, such as NVLink, are crucial for efficient inter-GPU communication, reducing latency and improving performance. NVLink provides direct, high-bandwidth connections between GPUs within a node, so running jobs within a single node helps improve training efficiency.

\section{Motivation}
\subsection{Existing Solutions}\label{AA}

\subsubsection{Serverless Computing }

In the early stages of LLM training, allocating GPU resources was a manual and complex process. Researchers and engineers had to specify the number of GPU cards required, considering factors like GPU memory capacity, model size, dataset size, batch size, and parallelism strategy. Different GPU types have varying memory capacities, and larger models require more GPU memory and computational power. The training dataset size and batch size also impact GPU memory needs. The parallelism strategy (data, model, or pipeline) affects the number of GPUs required and training efficiency. However, this manual approach often led to resource underutilization, over-provisioning, and extensive trial and error, which were time-consuming and resource-intensive. 


Recent work, such as ElasticFlow \cite{ElasticFlow}, has introduced automated methods for GPU allocation in large-scale training environments. Operating in homogeneous clusters, ElasticFlow uses admission control to dynamically determine and allocate the necessary number of GPUs for a task, scaling based on workload and job deadlines. However, ElasticFlow does not consider GPU memory capacity and heterogeneous resources, which can lead to inefficient use of resources. This limitation often results in a high rate of trial and error, reducing overall training efficiency. 

\subsubsection{Heterogeneous training}
Earlier LLM training efforts predominantly relied on homogeneous GPU clusters, where all GPUs had the same specifications. Frameworks such as ALPA \cite{Alpa}, POLLUX \cite{Pollux}, and Gandiva \cite{Gandiva} were designed to optimize training on these uniform clusters. These systems provided efficient parallelization and resource management within the constraints of identical hardware. However, this approach is relatively limited, as it does not leverage the diverse computational and memory capabilities of different GPU types available in modern data centers. The uniformity of the cluster restricts the flexibility and potential efficiency gains that could be achieved with a more diverse set of resources.


Recent work has started to address the challenges of training on heterogeneous GPU clusters. Metis \cite{Metis} automates the setup and deployment of parallelization strategies across different GPU types, dynamically determining the most efficient approach to maximize resource utilization and training speed. Sia \cite{Sia} focuses on optimizing job scheduling in heterogeneous GPU clusters, assigning tasks to the most suitable GPUs based on their computational and memory capabilities. Despite these advancements, both Metis and Sia face significant challenges. They overlook the vast search space required to estimate the optimal number of different GPU types needed for a given task, making it difficult to manually specify the correct number of GPUs. Additionally, the complexity of job scheduling in heterogeneous environments adds another layer of difficulty, as the varying capabilities of different GPUs can lead to suboptimal resource allocation and inefficient training processes.

\subsubsection{Our solution}
To address these issues, we propose Frenzy, a serverless system designed to automate heterogeneous training process. In Frenzy, users only need to submit their LLM training tasks without worrying about the specific GPU types or the number of GPUs required. The system automatically identifies the model parameters and the available resources in the heterogeneous cluster, performing fully automated model deployment and training.

\subsection{Challenges}
\subsubsection{Challenges in Automating GPU Card Number Decision}
Achieving automatic decision-making for the number of GPU cards required for LLM training is a highly complex problem.

Several key factors contribute to the complexity of training LLMs on heterogeneous GPU clusters. The model size and batch size significantly impact GPU memory requirements; larger models and batch sizes demand more memory, and failing to account for these can lead to out-of-memory errors. The choice of parallelism strategy (data, model, or pipeline) influences the number of GPUs needed, with each strategy having different memory and computational requirements. In a heterogeneous cluster, different GPUs have varying memory capacities, adding complexity as the system must consider the available GPU types to avoid memory issues. Managing dynamic memory usage (e.g., forward and backward passes, gradients) and static memory usage (e.g., model weights, optimizer states) is crucial for smooth training \cite{zero}. The interplay between these factors creates a complex search space, making it challenging to manually specify the number and types of GPUs. Addressing these challenges requires a sophisticated system that can dynamically adapt to the diverse conditions of a heterogeneous cluster, ensuring efficient LLM training. 

\subsubsection{Challenges in Low-overhead Scheduling for Heterogeneous GPU Clusters}



Once the required number of GPU cards for LLM training has been determined, the challenge shifts to scheduling these resources in a heterogeneous GPU cluster, which is a highly intricate and complex task. To ensure low scheduling overhead and efficient resource utilization, several critical factors must be addressed. These include carefully matching the memory requirements of the model to the varying memory capacities of different GPUs to prevent out-of-memory errors and ensure smooth training. Efficient scheduling also involves maximizing throughput by strategically placing tasks to optimize communication paths. Additionally, balancing the load across GPUs with different computational powers is crucial to avoid bottlenecks and ensure optimal performance. By addressing the interplay between memory capacity, communication bandwidth, and computational power, the scheduling system can minimize the complexity and overhead, ensuring that the training process is both efficient and effective.

\section{Design}
\begin{figure}
    \centering
    \includegraphics[width=1\linewidth]{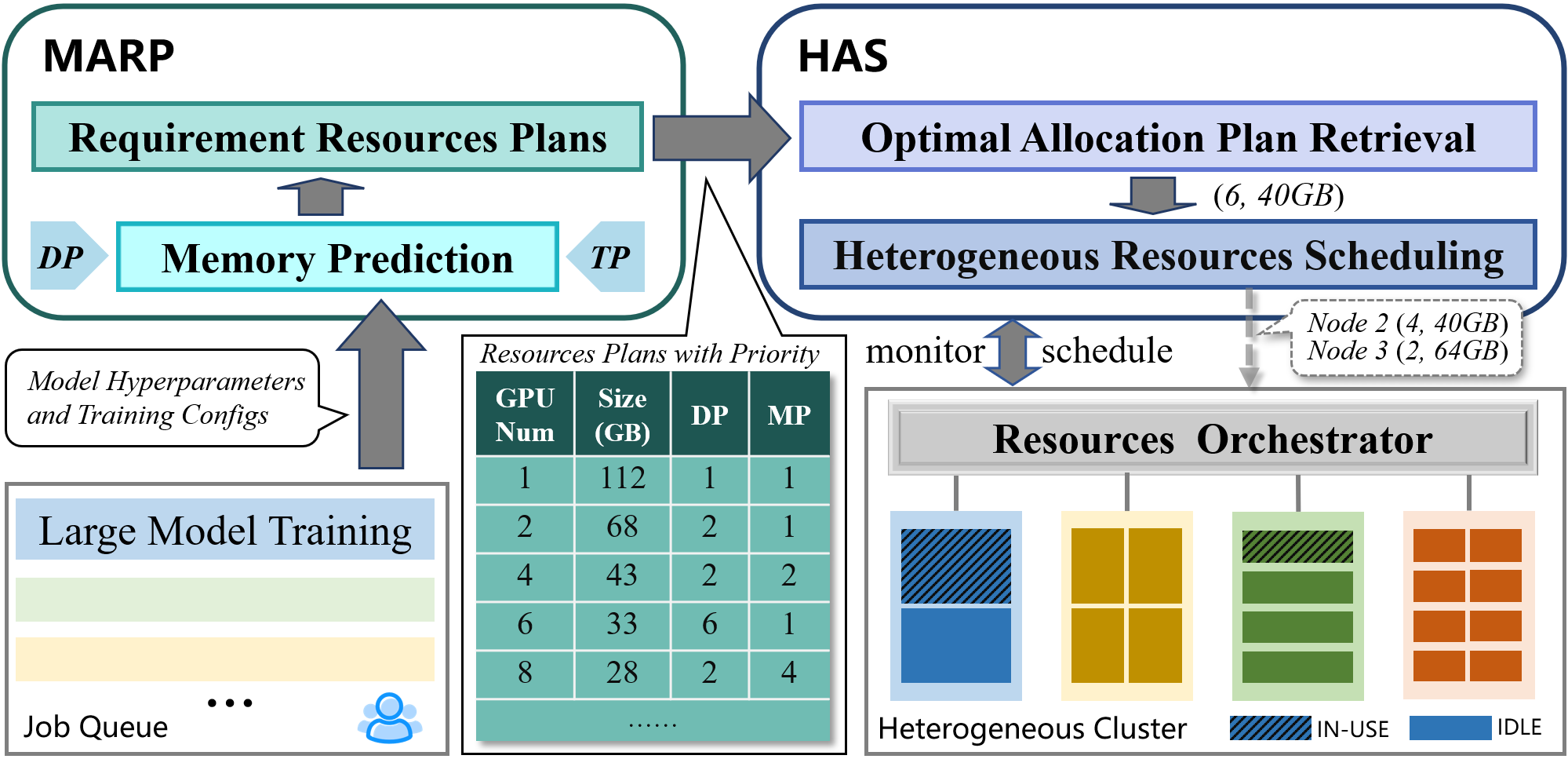}
    \caption{Frenzy overview. User submits a large model training job. \textbf{MARP} predicts the required training resources based on the model hyper-parameters and training configurations, combined with different data parallelism and tensor parallelism numbers, and outputs multiple required resources plans with priorities. \textbf{HAS} retrieves the optimal resource allocation plan among them and then schedule resources based on the heterogeneous GPU cluster.}
    \label{overview}
\end{figure}

To address the aforementioned two challenges, we have designed MARP (Memory-Aware Resource Predictor) and HAS (Heterogeneity-Aware Scheduler) accordingly. The overall architecture is illustrated in Figure \ref{overview}. 

Frenzy consists of three key components: MARP, HAS, and the Resource Orchestrator.

\begin{itemize}
    \item  MARP is designed to predict the required number of GPU cards for the training process. It takes into account the parameters of the LLM model submitted by the user, the batch size of the input data, and the memory capacities of different GPUs. By analyzing these inputs, MARP outputs the optimal number of each type of GPU needed to ensure efficient and feasible training.
    \item  HAS receives the output from MARP and integrates it with information about the available resources in the heterogeneous cluster. It also considers the estimated training time for the model on different GPU types. HAS then performs a sophisticated scheduling process to allocate resources in a way that maximizes efficiency and minimizes training time, ensuring that the training job is executed optimally across the available GPUs.
    \item The Resource Orchestrator manages and orchestrates the heterogeneous GPU resources. It dynamically records and aggregates available resources, and executes the allocation and release of these resources to ensure smooth and efficient training.
\end{itemize}
This integrated approach ensures that Frenzy can effectively handle the complexities of training large language models on heterogeneous GPU clusters.

\subsection{MARP}

\begin{figure}
    \centering
    \includegraphics[width=1\linewidth]{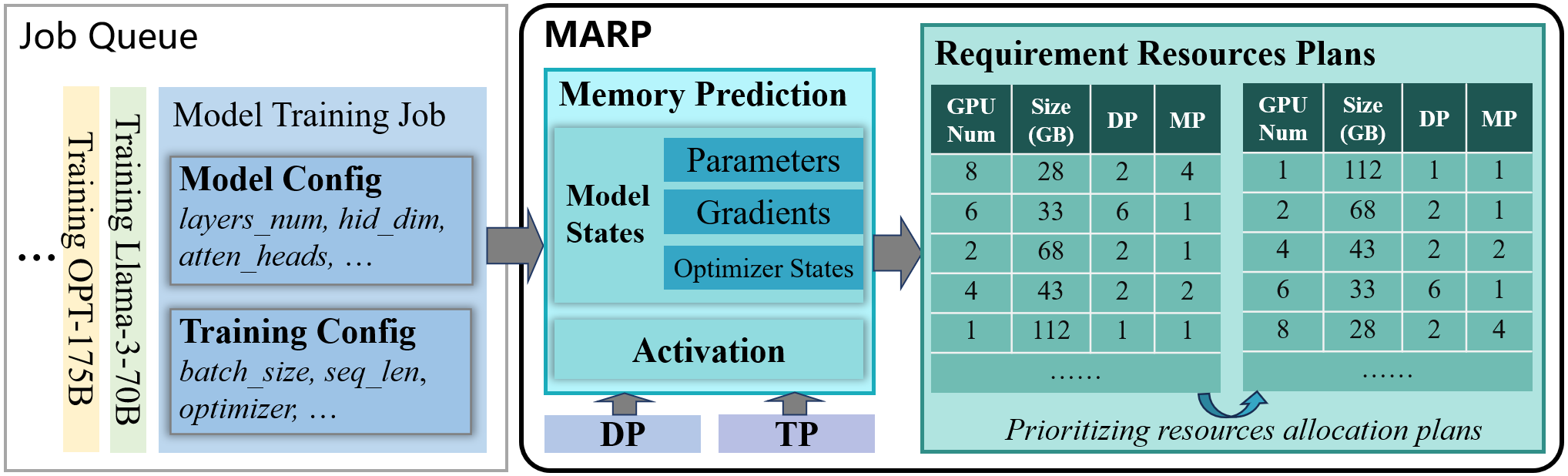}
    \caption{\textbf{MARP}. For a training job, MARP calculates the GPU memory that will be occupied by Model States and Activations during training, based on the model and training config, under different numbers of data parallelism and tensor parallelism. MARP adopts a priority ranking for the requirements training resources plans obtained from different parallel schemes.}
    \label{marp}
\end{figure}

The main idea behind MARP is to estimate the GPU memory usage during training based on the user-input model parameters and training configurations, and then determine the required number of GPU cards for different types of GPU. As can be seen in Figure \ref{marp}.

LLM, especially decoder-only models, have reached a high level of maturity in their structural design, typically consisting of multiple layers of self-attention mechanisms and MLPs. This standardized architecture simplifies model design and is crucial for optimizing resource utilization, allowing for more accurate prediction of peak GPU memory requirements during training. The parallelization strategy also significantly impacts GPU memory usage. Data parallelism\cite{zero}, which splits the global batch size across multiple GPUs, accelerates training but does not reduce individual GPU memory usage. Tensor parallelism\cite{megatron}, which distributes different parts of the model across multiple GPUs, is effective for large models by reducing the memory burden on individual GPUs but increases communication overhead. Pipeline parallelism improves computational efficiency by assigning different layers to different devices but does not reduce activation memory. Therefore, when predicting GPU memory usage, we primarily consider data parallelism and tensor parallelism. 

GPU memory mainly contains static and dynamic parameters. Static parameters include model parameters and optimizer states, while dynamic parameters are primarily activations. Assuming the weight parameter count is \textit{W} and adam optimizer \cite{adam} has been used, in mixed-precision training \cite{mix-prec}, the size of static parameters is 20\textit{W}\cite{Smith}. Through profiling LLM weight, we can approximate \textit{W}  using the following formula: 
\[ \textbf{\textit{W}}=\textit{V}\textit{h}+\textit{l}(12h^2+13\textit{h})\]
where \textit{V} is the vocabulary size, \textit{h} is the hidden layer size, and \textit{l} is the number of layers.

All static parameters can be split using tensor parallelism, meaning the size of the static parameters in every single GPU is \(\frac{20\textit{W}}{t}\) , where \textit{t} is the size of the tensor parallelism.
In mixed-precision training, previous work has shown that the dynamic parameter usage of activation is given by \cite{reduc-acti}:
\[activations=sbhl(10+\frac{24}{t}+\frac{5as}{ht})\]
where \textit{b} is micro batch size, \textit{s} is sequence length, \textit{a} is number of heads.

In real clusters, there are GPUs with different memory capacities. For each type of GPU, we need to ensure that the total sum of dynamic and static parameters generated during model training is less than the total GPU capacity. Typically, users specify the batch size \textit{B} of input data when training a model, and we need to decompose it into multiple micro batch size\textit{b} through data parallelism, i.e., \(b=\frac{B}{d}\), where 
\textit{d} is the size of data parallelism.
The specific memory estimation formula is as follows:
\[\frac{20\textit{W}}{t}+sBhl(\frac{10}{d}+\frac{24}{dt}+\frac{5as}{dht}) < GPU capacity\]
By adjusting the following parameters, we can meet the memory constraints: 

\begin{itemize}
    \item Size of data parallelism \textit{d}: Increasing the size of data parallelism can reduce the micro batch size on each GPU, thereby reducing the dynamic parameter usage.
    \item Size of tensor parallelism \textit{t}: Increasing the size of tensor parallelism can further reduce the static parameter usage on each GPU.
\end{itemize}
 
Then we can determine the required number of GPU cards \textit{N}, where \textit{N} = \textit{d} × \textit{t}, thereby realizing serverless computing. 

\subsection{HAS}
\begin{figure}
    \centering
    \includegraphics[width=1\linewidth]{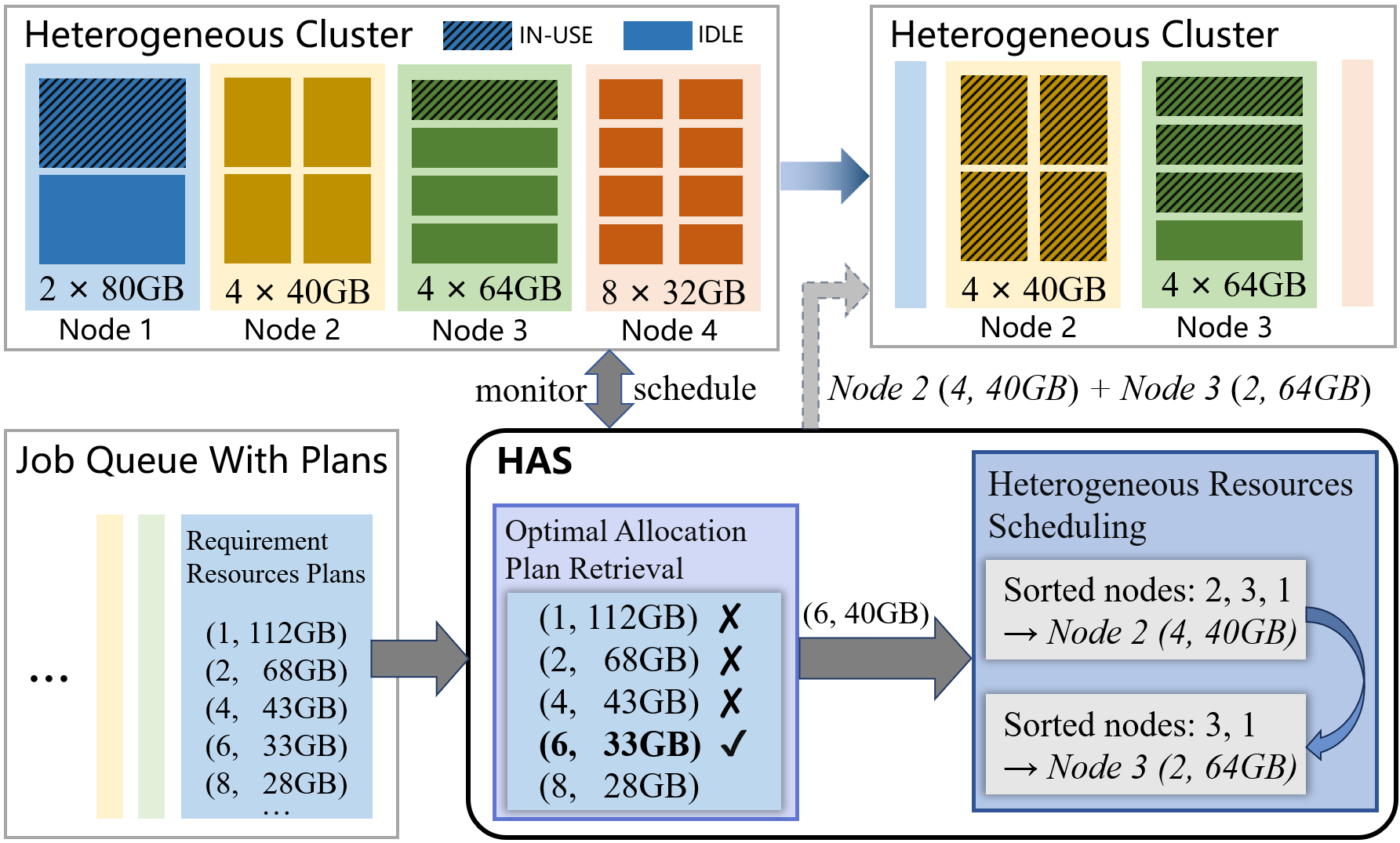}
    \caption{\textbf{HAS}. For the various resources allocation plans with priorities output by MARP, HAS conducts a sequential search based on the current resource status of the cluster to obtain the optimal resource plan that can be satisfied. Then HAS allocates resources for the job based on the heterogeneous GPU cluster based on this plan.}
    \label{fg.HAS}
\vspace{-3pt}
\end{figure}
For different training tasks, their requirements for the number of GPUs and memory size vary. Concurrently, it is common in reality that different nodes within a cluster possess various types of GPUs. Consequently, different scheduling strategies for the same task can significantly impact the utilization of cluster resources and the efficiency of job execution.

In the context of cluster resource management for training tasks, we utilize the notation $Node(n, s)$ to describe a specific node, where $n$ represents the number of remaining GPU cards on the node, and $s$ represents the size of each GPU. Correspondingly, a job is represented as $Job(n, s)$, with $n$ denoting the number of GPUs required by the job, and $s$ specifying the minimum size necessary for each GPU. For example, $Node(2, 40)$ refers to a node with two remaining 40GB GPU cards, and $Job(2, 32)$ signifies a task that requires 2 GPUs, each with at least 36GB of memory.
It is evident that for $Job(2, 32)$, allocating it to $Node(3, 40)$ is considered more efficient in terms of cluster resource utilization compared to assigning it to $Node(6, 80)$. For $Job(4, 35)$, it is more appropriate to schedule it on $Node(4, 40)$ rather than distributing it across four $Node(1, 40)$ units. This preference arises from the fact that the communication overhead between separate nodes can significantly prolong the runtime of the job.

In addition to the optimal resource allocation for a job possibly not being met by the existing resources in the cluster, such as a Job that could be satisfied by allocating resources of $(1, 45GB)$ or $(2, 35GB)$, but the cluster currently lacks available GPUs larger than 45GB, so it has to settle for allocating two 40GB cards, or even a less optimal resource allocation plan.

\begin{algorithm}
\caption{\textbf{Function} \textsc{HAS}(JobResourcePlans, Cluster)}
\label{alg.HAS}
\begin{algorithmic}[1] 
    \STATE Get available resources $avaRsc$ from $Cluster$
    \STATE $optimalPlan$ = None
    \FOR{$Plan$ in $JobRespourcePlans$}
        \STATE Get $reqNum$, $reqSz$ from $Plan$
        \STATE $ava$=sum(1 for gpu in $avaRsc$ if gpu.size $\ge reqSze$)
        \IF {$ava \ge reqNum$}
            \STATE $optimalPlan$ = $Plan$
            \STATE break
        \ENDIF
    \ENDFOR

    \STATE Get $reqNum$, $reqSz$ from $optimalPlan$
    \STATE $allocList$ = [ ]
    \WHILE{$reqNum > 0$}
        \STATE $fitSz$=min(g.size for g in $avaRsc$ if g.size $\ge reqSz$)
        \STATE $NLst$=[n for n in $cluster$ if n.gpusize $> fitSz$]
        \STATE $NLst$=sort($NLst$, key=$NLst.idleGPUs$, asc=$True$)
        \STATE $singleNodeMeetRequirement$ = $False$
        \FOR{each node $N$ in $NLst$}
            \IF{$N.idleGPUs \ge reqNum$}
                \STATE add $(N.id, reqNum)$ to $allocList$
                \STATE $N.idleGPUs$ = $N.idleGPUs - reqNum$
                \STATE $reqNumv$ = $0$
                \STATE $singleNodeMeetRequirement$ = $True$
                \STATE break
            \ENDIF
        \ENDFOR
        \IF{$singleNodeMeetRequirement$}
            \STATE break
        \ELSE
            \STATE $N$ = $NLst$[$-1$]
            \STATE add $(N.id,N.idleGPUs)$ to $allocList$
            \STATE $reqNum$ = $reqNum - N.idleGPUs$
            \STATE $NLst$[$-1$]$.idleGPUs$ = $0$
        \ENDIF
        \STATE update $avaRsc$
    \ENDWHILE
    \RETURN $allocList$
\end{algorithmic}
\end{algorithm}

Therefore, how to place jobs more reasonably on suitable nodes requires careful consideration. We have meticulously designed a heterogeneous-aware scheduler named HAS, which addresses the aforementioned issues. Figure \ref{fg.HAS} illustrates the resource scheduling process of HAS, which consists of two stages:

\begin{itemize}
    \item \textbf{Optimal Resource Allocation Plan Retrieval}: After a training job undergoes MARP calculation, it yields multiple resource allocation plans. The plans at the forefront indicate higher training efficiency for the job. HAS traverses these plans from top to bottom based on the available resources in the cluster until it finds a resource allocation plan that the cluster can satisfy at that moment. This plan represents the optimal solution for the job under the current available resources in the cluster. 
    \item \textbf{Heterogeneous Resource Scheduling}: Upon obtaining the plan, HAS proceeds with resource scheduling based on the available resources in the heterogeneous cluster. Specifically, HAS employs a Best-fit strategy, allocating nodes in the cluster that exactly meet the resource requirements to the job. If there is no single node that exactly meets the above resource requirements, it allocates the node with the least remaining resources among those that can meet the requirements. If no single node can meet the resource requirements, it adopts a greedy strategy, allocating the node with the most remaining resources to the job, calculating the remaining required resources, and repeating the above steps. 
\end{itemize}

Algorithm \ref{alg.HAS} embodies the whole idea of HAS.

\section{Evaluation}
To validate the effectiveness of Frenzy, we conducted multiple experiments, including tests to measure the average job completion time, scheduling overhead, and memory prediction accuracy.

\subsection{Setup}\label{AAA}
\paragraph{Cluster configuration and simulator}

(1) Real tested: We conducted LLM training task scheduling experiments using Ray on physical cluster with following setups: 1 node with 2 $\times$ A100 40G GPUs(head node) interconnected with PCIe, 1 node with 1 $\times$ A100 40G GPUs, 1 node with 4 $\times$A800 80G GPUs interconnected with NVlink, and 2 nodes with 2 $\times$A100 80G GPUs interconnected with PCIe, total of 5 nodes with 3 different types of GPUs.
(2) Simulator: We established a simulation environment using the PAI platform simulator, developed and validated by Alibaba Cloud \cite{PAI}, and used it to verify the effectiveness of Frenzy. We set up the same experimental configuration as Sia \cite{Sia}, which includes: 3 nodes with 8 NVIDIA 2080Ti GPUs each, 2 nodes with 8  A100 40G GPUs each, and 1 node with 4 RTX6000 GPUs. 

\paragraph{Workload and traces  }

\textit{Philly} \cite{LagSca-Clus-Ana} is a dataset derived from over 100,000 jobs executed over a two-month period in a multi-tenant cluster with multiple GPU types at Microsoft. This dataset provides valuable insights into the real-world usage patterns and challenges of managing large-scale, heterogeneous GPU clusters. 
\textit{Helio} \cite{Helio} is a real-world workload dataset from SenseTime, consisting of data from four clusters and including over 3.3M tasks. Compared to\textit{ Philly}, \textit{Helio} requires more GPUs and has longer runtime durations. 
We also used \textit{NewWorkload}, consisting of GPT-2 \cite{gpt2} and BERT \cite{bert} models with different sizes and various batch sizes, to create 30 and 60 LLM task queues, respectively. 

\paragraph{Baseline}
We have two counterparts, including Sia and \textit{Opportunistic Scheduling}.
Sia \cite{Sia} optimizes deep learning (DL) scheduling in heterogeneous clusters, enabling adaptive scheduling for tasks with user-speciffed numbers of GPUs.
\textit{Opportunistic Scheduling}\cite{Lyra} denotes the version with opportunistic scheduling strategy, which always prioritizes nodes with higher computational power in heterogeneous cluster scheduling. It follows a first-come, first-served (FCFS) policy, greedily allocating idle resources to newly submitted tasks. 

\subsection{General result}\label{FAT}
\begin{figure}
    \centering
    \begin{subfigure}[b]{0.37\linewidth}
        \centering
        \includegraphics[width=\linewidth]{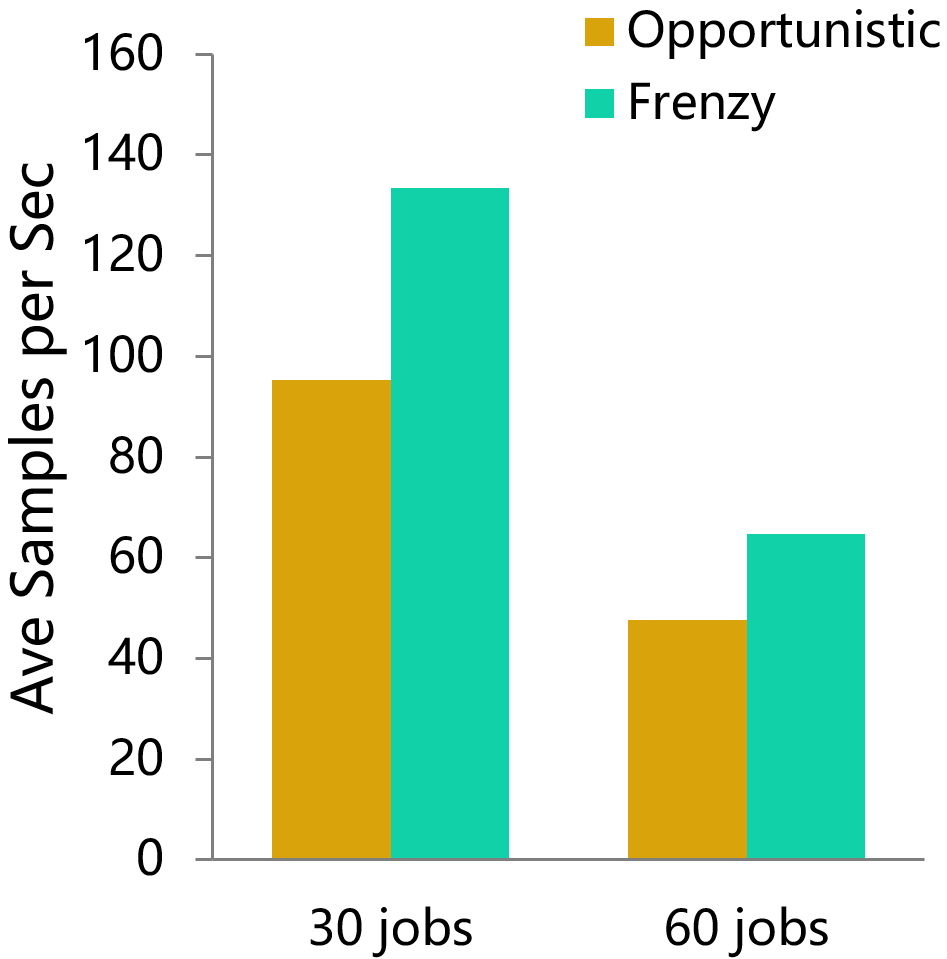}
        \caption{}
    \end{subfigure}
    \hfill 
    \begin{subfigure}[b]{0.59\linewidth}
        \centering
        \includegraphics[width=\linewidth]{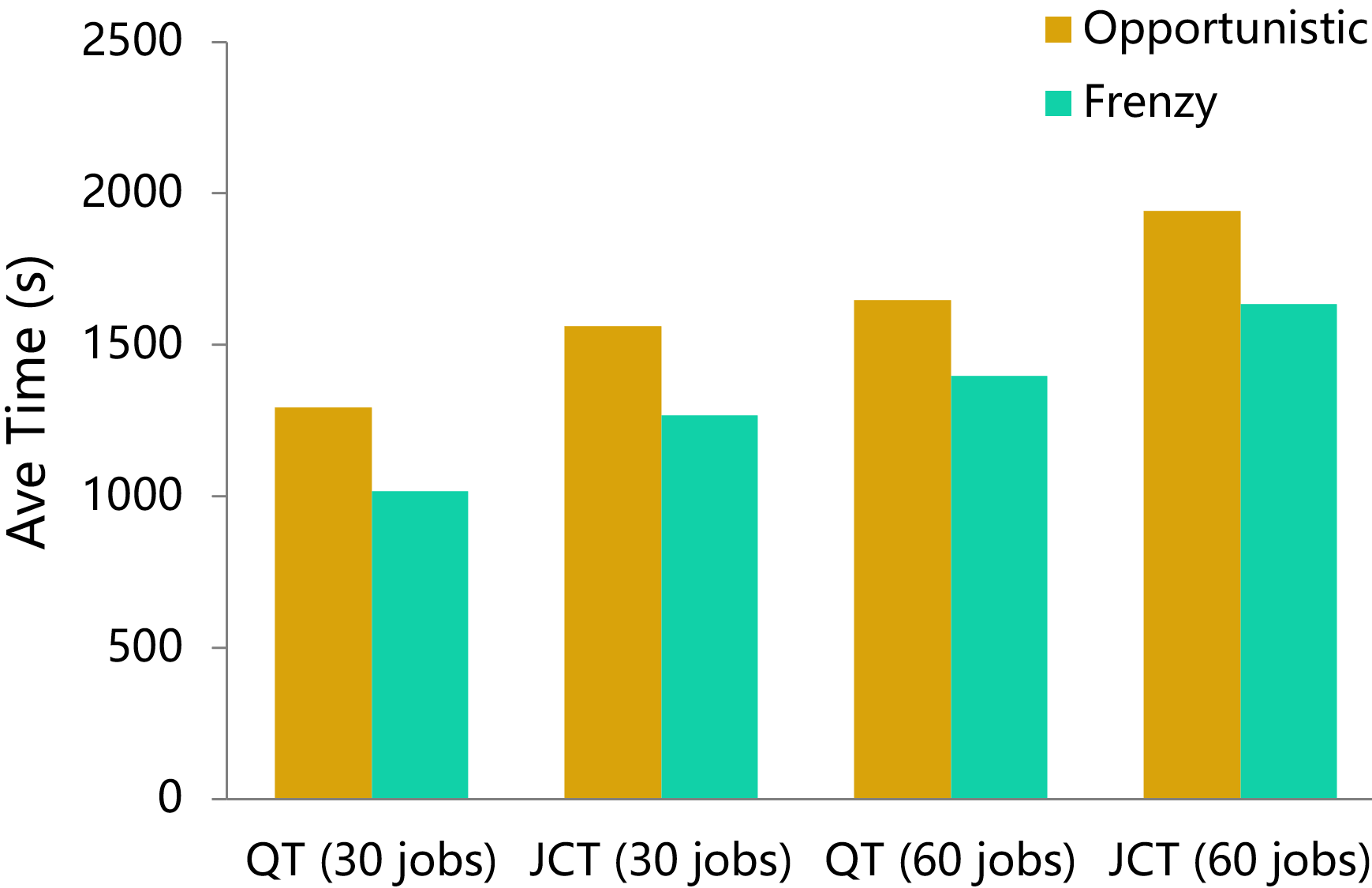}
        \caption{}
    \end{subfigure}
    \caption{Comparison with opportunistic scheduling. QT means Queue time, and JCT means Job complete time.}
    \label{comparison_realtest}
\end{figure}

We tested the average number of samples completed per job per second and the average job completion time for the task queues based on \textit{NewWorkload}. The results are shown in Figure \ref{comparison_realtest}. From Figure \ref{comparison_realtest}(a), it is evident that for workloads of 30 and 60 tasks, the average number of samples completed per second per model increased by 29\% and 27\% compared to \textit{opportunistic scheduling}, respectively. Figure \ref{comparison_realtest}(b) shows that for workloads of 30 and 60 tasks, the average queue time and average job completion time decreased by 13.7\% and 18.1\% for the 30-task workload, and by 15.2\% and 15.8\% for the 60-task workload. 

\begin{figure}
    \centering
    \begin{subfigure}[b]{0.67\linewidth}
        \centering
        \includegraphics[width=\linewidth]{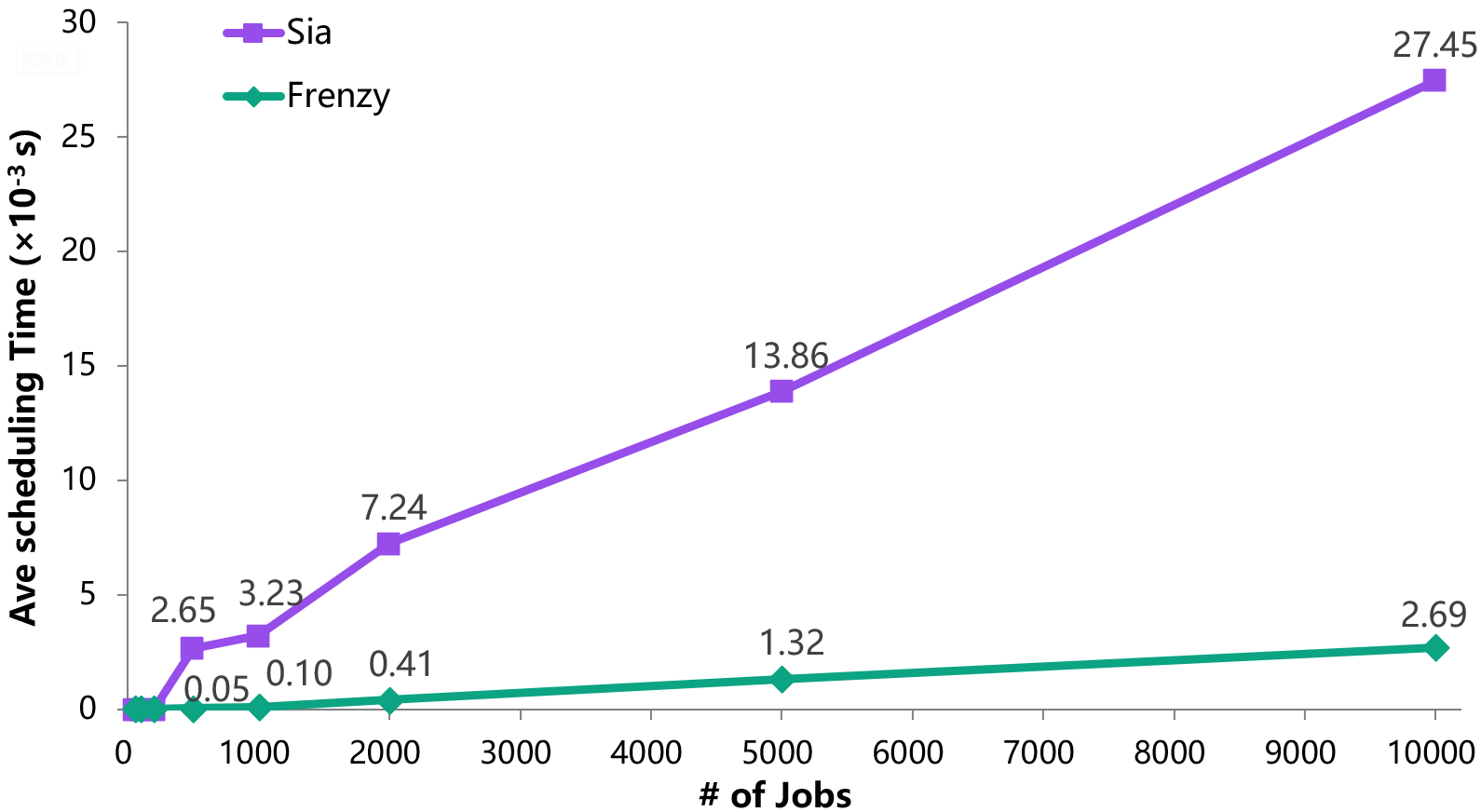}
        \caption{}
    \end{subfigure}
    \hfill 
    \begin{subfigure}[b]{0.31\linewidth}
        \centering
        \includegraphics[width=\linewidth]{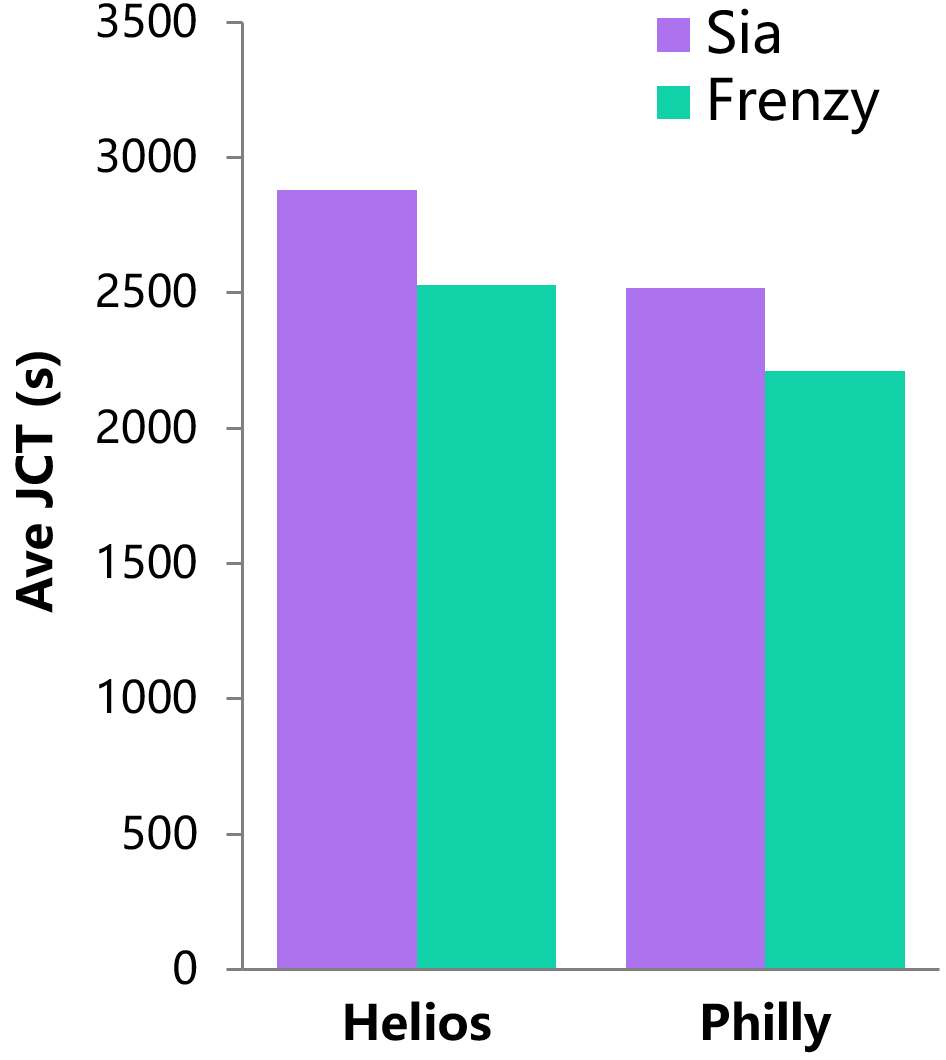}
        \caption{}
    \end{subfigure}
    \caption{Frenzy scheduling result compared with Sia}
    \label{com-sia}
\end{figure}

Then, we compared our work with state-of-the-art (SOTA) approaches using PAI simulator \cite{PAI}. 
First, we compared the scheduling overhead of Frenzy with Sia. As shown in Figure \ref{com-sia}(a), Sia's scheduling algorithm exhibits extremely rapidly increasing overhead as the number of tasks grows, resulting in significantly much higher scheduling costs compared to Frenzy. Additionally, we compared the average task completion time of Frenzy and Sia based on \textit{Helios} \cite{Helio} and \textit{Philly} \cite{LagSca-Clus-Ana} traces.
The comparison results are shown in Figure \ref{com-sia}(b). Compared to Sia, our average task completion time was reduced by approximately 12\% both on \textit{Helios} and \textit{Philly}.

Compared to SOTA works, Frenzy achieves superior performance for two reasons: First, MARP provides more reasonable resource requests for each LLM training task. Second, HAS enables low-overhead and efficient scheduling for heterogeneous GPUs. 

\subsection{Evaluation for accuracy of memory prediction by Frenzy}
\begin{figure}
    \centering
    \includegraphics[width=1\linewidth]{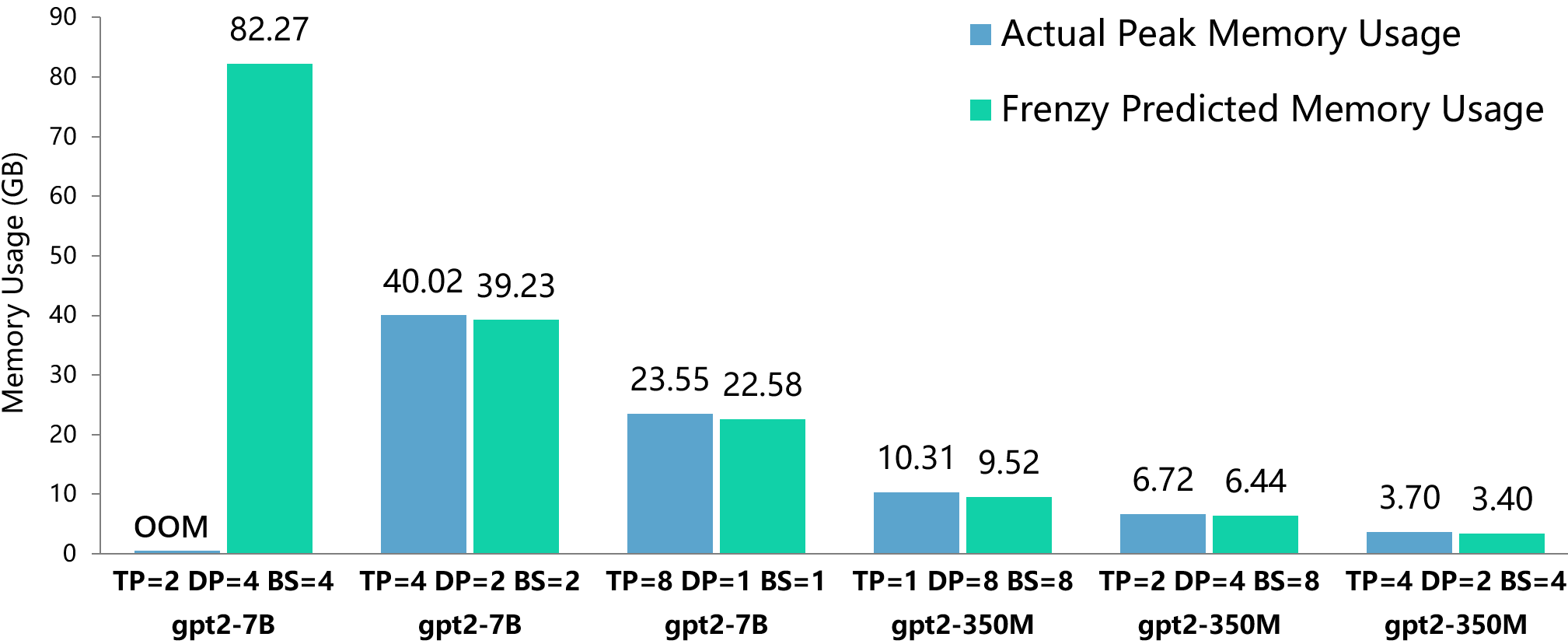}
    \caption{Frenzy memory prediction compared with reality}
    \label{memory-comparison}
\end{figure}

We conducted experiments on GPT2-7B and GPT2-350M for real tested. In these experiments, we predicted the peak memory usage of large language models (LLMs) under different parallelization strategies and batch sizes. The experimental results are shown in Figure \ref{memory-comparison}. We can see that the peak memory prediction accuracy of MARP ranges from 92\% to 98\% across different parameter configurations, which plays a decisive role in determining the number of GPUs required for different categories. For example, when training the GPT2-7B model with a batch size of 2, 8 cards of A100 GPUs are needed for model training, and the GPU utilization is relatively highest when tensor parallelism is 4 and data parallelism is 2.  

\section{Conclusion}

Frenzy is the first system to integrate serverless computing into heterogeneous clusters, significantly reducing the burden on developers. By accurately predicting peak memory usage during LLM training, Frenzy estimates the required heterogeneous resources, enabling low-overhead and efficient heterogeneity-aware scheduling. This allows users to submit tasks without concern for complex underlying hardware configurations, as the system automatically handles resource requests and deployment. Moreover, Frenzy demonstrates superior performance with a lower average job completion time, precise memory prediction, and reduced scheduling overhead compared to existing solutions. 

\IEEEtriggeratref{13}

\vspace{12pt}

\end{document}